\documentclass[10pt,journal]{IEEEtran}
\usepackage{fixltx2e}
\usepackage{amsmath,amsthm}
\usepackage{url}
\usepackage{romannum}
\usepackage{import}
\usepackage{pgfplots}
\usetikzlibrary{pgfplots.groupplots}
\usepackage{caption}
\usepackage{paralist}
\usepackage{enumitem}
\usetikzlibrary{spy}

\newcommand{\cov}[1]{\text{cov}\left\{#1\right\}}
\newcommand{\trace}[1]{\text{tr}\left\{#1\right\}}
\newcommand{\expect}[1]{\text{E}\left\{ #1 \right\}}
\newcommand{\vecterize}[1]{\text{vect}\left\{ #1 \right\}}
\newcommand{\etal}{\emph{et al.~}}

\newtheorem{Remark}{Remark}

\usepackage{amsfonts}
\usepackage{amssymb}
\usepackage{mathrsfs}
\usepackage{amsmath}
\usepackage{amssymb}
\usepackage{dsfont}
\usepackage{url}

\usepackage{xargs}
\usepackage{color}

\definecolor{tcolor}{RGB}{0,60,104}
\usepackage{graphicx}

\newcommand*\vect[1]{\begin{bmatrix}#1\end{bmatrix}}
\newcommand*\tvect[1]{\big[#1\big]^T}

\newcommand{\tp}{\mathrm{T}}

\newcommand{\rvec}[1]{\ensuremath{{\boldsymbol{{#1}}}}}
\newcommand{\mat}[1]{{\ensuremath{{\mathbf{#1}}}}}

\newcommand{\IR}{\NewR}

\newcommand{\NewR}{\ensuremath{\mathds{R}}}
\def\Eq#1{\eqref{#1}}
\def\Sec#1{Section~\ref{#1}}

\def\Eq#1{\eqref{#1}}
\def\Sec#1{Section~\ref{#1}}

\DeclareMathOperator{\diag}{diag}

\newcommand{\Fig}[1]{Fig.~\ref{#1}}

\begin{document}
\title{Tracking the Orientation and Axes Lengths of an Elliptical Extended Object}

\author{Shishan~Yang and
		Marcus~Baum
		\IEEEcompsocitemizethanks{\IEEEcompsocthanksitem S. Yang and M. Baum are with the Institute of Computer Science,
								 University of Goettingen, Goettingen 37077, Germany
								 (e-mail: shishan.yang@cs.uni-goettingen.de; marcus.baum@cs.uni-goettingen.de).\protect\\
									}
}

\maketitle

\begin{abstract}
Extended object tracking considers the simultaneous estimation of the kinematic state and the shape parameters of a moving object based  on a varying number of noisy detections. 
A main challenge in extended object tracking is the  nonlinearity and high-dimensionality of the estimation problem.
This work presents compact closed-form expressions for a recursive Kalman filter that explicitly estimates the orientation and axes lengths of an extended object based on detections that are scattered over the object surface (according to a Gaussian distribution). 
Existing approaches are either based on Monte Carlo approximations or do not allow for explicitly maintaining all ellipse parameters. The performance of the novel approach is demonstrated with respect to the state-of-the-art by means of simulations.
\end{abstract}

\IEEEpeerreviewmaketitle
\section{Introduction}
\label{sec:intro}

The objective of extended object tracking is to simultaneously determine both the kinematic state and the shape parameters of a moving object.
With the development of novel near-field and high-resolution sensors, extended object tracking is becoming increasingly important in many applications such as autonomous driving \cite{KunzEtAl:2015,HirscherSRD:2016} and  maritime surveillance \cite{Vivone2016}. Recent overviews of extended object tracking methods and applications are given in \cite{Granstroem2017} and \cite{Mihaylova2014}.

 Most sensors for extended object tracking, e.g., LiDAR or radar devices, provide a varying number of spatially distributed detections (measurements) per  scan from the object.
 Depending on the specific sensor and target, different scattering patterns can be distinguished. For example, in two-dimensional space,  measurements can be scattered on the surface of the object, or on the boundary of the object.

\begin{figure}
\def\svgwidth{0.8\columnwidth}
\footnotesize
\centering
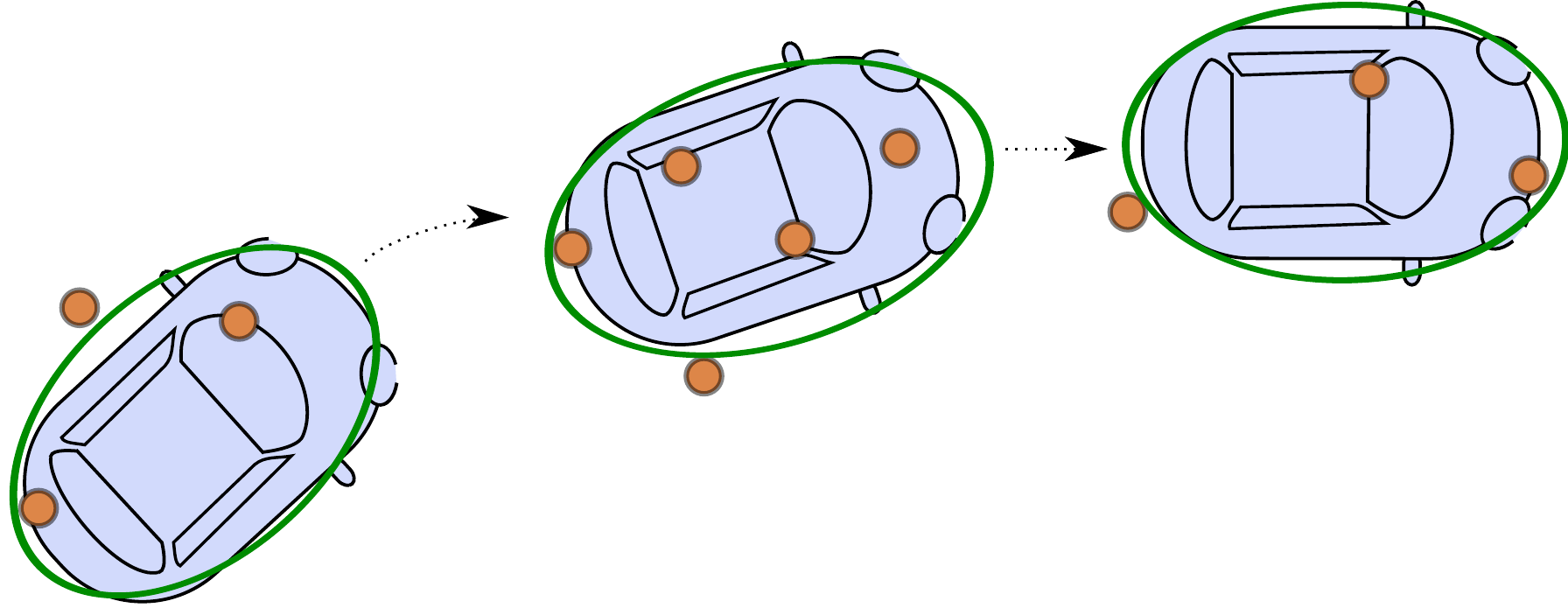
\caption{Illustration of the extended object tracking problem. Multiple spatially distributed measurements are received from the target object and the objective is to determine an elliptic shape approximation in addition to the kinematic target state.}
\label{fig:EOT}
\end{figure}

In case of  spatially dense measurements, it might be possible to extract detailed shape information from the object. 
For example, star-convex shape approximations as in \cite{Fusion11_Baum-RHM,Kaulbersch2017,J_TAES_Baum_RHM,Wahlstroem2015,Oezkan2016} are widely-used for this purpose. 
In scenarios with high measurement noise and a relatively low number of measurements from the object, it  is common to approximate the object shape with an ellipse (see Fig.~\ref{fig:EOT}).                                                                                                                        The random matrix approaches \cite{Koch2008,Feldmann2010,Orguner2011,Lan2012b} pioneered by  Koch \cite{Koch2008} can be seen as the state-of-the-art  for estimating elliptic shape approximations in case of surface scattering. 
 By means of representing the shape estimate and its uncertainty with an Inverse-Wishart density,
 it is possible to derive compact closed-form expression for a Bayesian measurement update. 
 The Inverse Wishart density is defined on symmetric positive-definite (SPD) matrices, where it is specified  by 
 \begin{itemize}
 \item a $d \times d$  SPD scale matrix  $\mat{V}$ and the 
 \item  scalar degree of freedom $v \in \IR$.
 \end{itemize}
 For $d=2$, the SPD scale matrix $\mat{V}$ can be interpreted as a two-dimensional elliptic shape estimate  with an uncertainty that is specified by the scalar $v$.
 An advantage of this representation is that the ellipse shape (including orientation, and semi-axes lengths) is uniquely defined by a single  scale matrix. However, the uncertainty of the complete ellipse shape  is encoded in a single one-dimensional value $v$. For this reason, it is not possible to distinguish explicitly between the uncertainty of the semi-axes and the orientation, which is often necessary in practical applications.

 \subsection{Contribution}
The main contribution of this work is a novel elliptic shape tracking method that explicitly maintains an
\begin{itemize}
 \item estimate for the orientation and semi-axes lengths, i.e., a three-dimensional vector, and
 \item the $3 \times 3$ joint covariance of the shape estimate.
\end{itemize}
 By this means, it becomes possible to  explicitly model the temporal evolution of  individual shape parameters and their interdependencies, which is highly relevant for numerous practical applications. For example, it can be directly modeled that the semi-axes are fixed (and unknown) but the orientation varies.

We derive compact closed-form expressions for a recursive update of the kinematic state  and the aforementioned shape parameters plus the respective covariance matrices.
Due to the high degree of nonlinearity of the problem,  a na\"ive application of standard estimation techniques such as the Extended Kalman Filter (EKF), Unscented Kalman Filter (UKF) or Second-Order Extended Kalman Filter (SOEKF)  \cite{Henriksen1982, Roth2011} result in unsatisfactory estimation results. In fact, we have even shown in our previous work \cite{Fusion12_Baum} that the  Linear Minimum Mean Squared Estimator (LMMSE) is already for the special case of axis-aligned ellipses inconsistent.

The key components that lead to the compact closed-form expressions are the followings:
\begin{compactenum}
\item[\textbf{C1}] An explicit measurement equation (corrupted by multiplicative noise) is formed that relates a measurement to the kinematic state and shape parameters
\item[\textbf{C2}] The kinematic state  and the shape parameters are decoupled, i.e., treated independently (as in the random matrix approach)
\item[\textbf{C3}] The kinematic state estimate is updated using the actual measurement. However, the shape parameters are updated with a pseudo-measurement constructed from the actual measurement
\item[\textbf{C4}] As the measurement equation for the kinematic parameters involves multiplicative noise,  a linearization is performed for the kinematic parameters, but the multiplicative noise is kept as a random variable for the moment calculation.  
\item[\textbf{C5}] A standard linearization of the measurement equation for the shape parameters does not yield a feasible estimator due to the high nonlinearities. For this reason, we 
derive a problem-tailored second-order approximation. In order to avoid the complex calculation of Hessian matrices, we exploit that the first two moments of the pseudo-measurement can be directly derived from the covariance matrix of the actual measurement.   
\end{compactenum}

This article is based on the two conference papers \cite{Yang2016,Yang2017_ICASSP}.
  Early ideas about the use of  a multiplicative noise term to model a  spatial distribution were discussed in  \cite{Fusion12_Baum}.
In   \cite{Yang2016}, we introduce a variant of the Second Order Extended Kalman filter (SOEKF)   for  estimating the orientation and semi-axes lengths of an ellipse.  Unfortunately, it involves complex calculations of several Hessian matrices.
In \cite{Yang2017_ICASSP}, we develop  a method that works completely without Hessian matrices.
To track an unknown number of extended objects,  an implementation that combines \cite{Yang2017_ICASSP} and Probability Density Hypotheses (PHD) filter is presented in \cite{Teich2017,Yang2019}.
The method introduced in this work improves  over \cite{Yang2017_ICASSP} by a more precise approximation of the  covariance of the predicted measurement. 
Furthermore, a much more detailed evaluation and comparison is provided. 

The shape modeling with multiplicative noise in (\textbf{C1}) is called \emph{Multiplicative Error Model (MEM)}. The  approximations (\textbf{C1})-(\textbf{C5})
are the key to a Kalman filter-based update. For this reason,
the new method is called \emph{MEM-EKF*}. EKF stands for Extended Kalman Filter (EKF) and the ``*'' emphasizes that (\textbf{C1})-(\textbf{C5}) are problem tailored linearization techniques and moment approximations. 
\subsection{Related Work}
Related work exists in the context of random matrix approaches,   random hypersurface models, and particle filtering, and optimization-based  approaches.

In \cite{Granstroem2013a}, an alternative prediction for the random matrix approach is derived that allows for kinematic state dependent predictions.
The model from Lan \etal  \cite{Lan2016b} can capture orientation changes using a  rotation matrix  and and isotropic scaling. However, both methods \cite{Granstroem2013a,Lan2016b} still work with Inverse Wishart densities, i.e., do not allow for explicitly maintaining the uncertainty of individual shape parameters.
The work \cite{Degerman2011} assumes the principal components of the measurements to be Gaussian and \cite{li2014ellipse} independently estimates the orientation and semi-axes lengths based on fitting ideas, i.e., both methods \cite{Degerman2011,li2014ellipse} are not based on the common spatial distribution model.

The random hypersurface approach \cite{J_TAES_Baum_RHM,SDF10_Baum} also allows to estimate elliptical shapes. However, the method discussed in \cite{J_TAES_Baum_RHM,SDF10_Baum}  uses the Cholesky decomposition of the shape matrix as a state vector, which has no intuitive meaning. Furthermore, the standard update in the random hypersurface approach requires a point estimate for the angle from the center to the measurement source, which is a poor approximation in case of  high measurement noise.
The roots of the proposed method here lie in the random hyperface approach, however, while the original random hypersurface approach uses a one-dimensional scaling factor, we here use a two-dimensional scaling.

A recent overview of particle filter methods for extended object and group tracking is given in \cite{Mihaylova2014}.
Based on Rao-Blackwellization, the random matrix approach has been combined with  particle filtering techniques \cite{vivone2017multiple}.
In \cite{Angelova2013},  a convolution particle filter is developed for tracking elliptical shaped extended objects.
Based on a hierarchical point process  and  particle-based approximations,    
multiple elliptic targets are tracked in \cite{swain2012phd} within the PHD filter framework.
Particle-based methods are also widely-used for group object tracking, e.g., \cite{septier2009tracking} considers a virtual leader model for groups and  a Markov Chain Monte Carlo (MCMC) method for approximating the posterior density.

A further class of methods optimizes over an entire batch of measurements in order to determine an estimate for the target object and its shape.
In \cite{Wieneke2012}, the Probabilistic Multi-Hypothesis Tracking (PMHT) in combination with the random matrix model is used to track multiple extended targets. 
In \cite{Kaulbersch2017}, a PMHT approach for estimating star-convex shape approximations is developed. In 
\cite{Brosseit2017}, a problem-tailored probability density function is introduced in order to determine the maximum likelihood estimate of elliptical target objects.

 \subsection{Structure}
 This article is structured as follows:
The next section introduces the basic models that are used for tracking an elliptical shape approximation of  a single extended object.
The following \Sec{sec:estimation} derives the compact closed-form expressions for a recursive measurement and time update.
A detailed evaluation of the proposed method is provided in \Sec{sec:eval}. Subsequently, this article is concluded in \Sec{sec:conclusion}.

\section{Modeling an Elliptical Extended Object}\label{sec:prob}

This section introduces the shape parameterization, measurement model, and process model for a single extended object whose shape is approximated as an ellipse.
\subsection{Parameterization}
The kinematic state of the object at time step $k$
\begin{equation}
\rvec{r}_k = \tvect{\rvec{m}_k^T, \rvec{\dot{m}}_k^T, \ldots}
\end{equation}
consists of the center $\rvec{m}_k \in \IR^2$, velocity $\rvec{\dot{m}}_k$,  and possible further quantities such as acceleration. 
As motivated in the introduction, the elliptical shape parameters at time $k$
  \begin{equation} \label{eq:param}
		\rvec{p}_k = \tvect{\alpha_k, l_{k,1}, l_{k,2} }\in \IR^3    ,
   \end{equation}
contains the 
\begin{compactitem}
\item angle $\alpha_k$, which indicates the counterclockwise angle of rotation from the $x$-axis, and the 
\item  semi-axes lengths $l_{k,1}$ and $l_{k,2}$.
\end{compactitem}
Although this ellipse parameterization is not unique, 
 it is quite common in tracking applications (for other measurement models), see for example \cite{Granstroem2011}.
The parameterization of the object extent as an SPD matrix $\mat{V}$ in the random matrix approaches is unique. 
However, the scalar parameter $v$ for the uncertainty does not allow to distinguish between the uncertainties of the orientation and the semi-axes lengths. A conversion from a covariance matrix of $\rvec{p}_k$ to scalar $v$   induces a loss of information.

\begin{figure}
\hspace{.7in}
\def\svgwidth{.6\columnwidth}
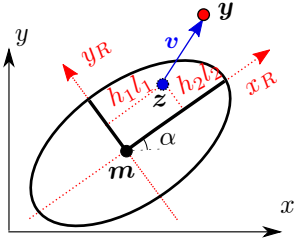
\caption{
	An illustration of our parameterization and measurement model. 
	We omit the time index $k$ and measurement index $i$ in this figure.
	The location of the object is $\rvec{m} = \tvect{m_{1},m_{2}}$.
	The object shape is denoted as $\rvec{p} = \tvect{\alpha, l_{1}, l_{2} }$.
 	The measurement $\rvec{y}$ is measurement source $\rvec{z}$ corrupted with measurement noise $\rvec{v}$.
	The measurement source  $\rvec{z}$ is related to $\rvec{p}$ using multiplicative noise $\rvec{h}=\tvect{h_1,h_2}$.
	By anticlockwise rotating coordinates system $x$-$y$ through an angle of $\alpha$,  we have the depicted ellipse axes-aligned in reference frame $x_R$-$y_R$.}
\label{fig:ellipse_par}
\end{figure}

\subsection{Measurement Model}

We adopt the widely-used spatial distribution model \cite{G2005,Gilholm2005} for modeling the object extent. 
The extended object gives rise to a varying number of independent two-dimensional Cartesian detections 
$$\mathcal{Y}_k=\{\rvec{y}_k^{(i)}\}_{i=1}^{n_k} $$ in each time step $k$. 
Each individual measurement (detection) $\rvec{y}_k^{(i)}$  originates from a measurement source $\rvec{z}_k^{(i)}$, which is corrupted by an additive Gaussian measurement noise $\rvec{v}_k^{(i)}$ with covariance of $\mat{C}^{\rvec{v}}$. 

Each measurement source $\rvec{z}_k^{(i)}$ lies on the object extent and follows a (uniform) spatial distribution.
Similar to the random matrix approach (e.g., \cite{Feldmann2010}), we approximate the uniform spatial distribution with a Gaussian spatial distribution. 

A key step to the proposed method -- see (\textbf{C1}) in the introduction -- is the formulation of an explicit measurement equation, which relates a measurement source and the object state with the help of  a multiplicative error term $\rvec{h}$. 
Consider an axis-aligned ellipse that lies in the origin and its semi-axes lengths are $l_1$ and $l_2$. Any point $\rvec{z}^{(i)}$ that lies on the ellipse can be written as
\begin{equation}\label{eq:axes_align}
  \rvec{z}^{(i)}  =   \vect{l_1&0\\0&l_2}\underbrace{\vect{h_1^{(i)}\\h_2^{(i)}}}_{:=\rvec{h}^{(i)}}\enspace.
\end{equation} 
To describe elliptical distributed measurement sources, we assume that the random variable $\rvec{h}^{(i)}$ is zero mean and  distributed on a unit circle. 
Note that if $h_1^{(i)}$ and $h_2^{(i)}$  would be independent and uniformly distributed on  the interval $[-1,1]$, we would model a rectangular spatial distribution with length  $2\cdot l_1$ and width $2 \cdot l_2$.

For an ellipse with orientation $\alpha$ and  center $\rvec{m}$ (see Fig.~\ref{fig:ellipse_par}), a rotation and translation transformation of \eqref{eq:axes_align} gives us
\begin{equation}\label{eq:non_axes_align}
  \rvec{z}^{(i)}  =  \rvec{m}+\underbrace{\vect{\cos{\alpha} & -\sin{\alpha}\\\sin{\alpha} & \cos{\alpha}}\vect{l_1&0\\0&l_2}}_{:=\mat{S}}\vect{h_1^{(i)}\\h_2^{(i)}}\enspace,
\end{equation} 
where  $\mat{S}$ specifies the orientation and size of the extended object.
  Incorporating the time index and sensor noise in \eqref{eq:non_axes_align} results in the measurement equation 
  \begin{equation}\label{eq:meas_kinematic}
  \rvec{y}_{k}^{(i)}=\mat{H}\rvec{r}_k + \mat{S}_k\rvec{h}_{k}^{(i)} +\rvec{v}_k^{(i)}\enspace,
\end{equation}
where  $\mat{H}=\vect{\mat{I}_2 & \mat{0}}$  picks the object location out of the kinematic state.

In the same way as \cite{Feldmann2010},  we assume  $\rvec{h}_k^{(i)}\sim \mathcal{N}(\mat{0},\mat{C}^{\rvec{h}})$ with
\begin{equation}
\mat{C}^{\rvec{h}} = \frac{1}{4} \mat{I}_2
\end{equation} in order to match the covariance of an elliptical uniform distribution.

\begin{Remark}
By assuming additive Gaussian measurement noise, the  measurement likelihood becomes
\begin{equation}\label{eq:MEMlikelihood}
p(\rvec{y}_k^{(i)}|\rvec{r}_k,\rvec{p}_k)=\mathcal{N}(\rvec{y}_k^{(i)};\mat{H}\rvec{r}_k,\mat{S}_k\mat{C}^{\rvec{h}}\mat{S}_k^{\tp}+\mat{C}^{\rvec{v}})\enspace.
\end{equation}
We would like to note that the measurement likelihood \eqref{eq:MEMlikelihood} is equivalent to the likelihood used in the random matrix approach \cite{Feldmann2010}, which is
\begin{equation}\label{eq:RMMlikelihood}
p(\rvec{y}_k^{(i)}|\rvec{r}_k,\mat{X}_k)=\mathcal{N}(\rvec{y}_k^{(i)};\mat{H}\rvec{r}_k,z\mat{X}_k+\mat{C}^{\rvec{v}})\enspace.
\end{equation} 
As $\mat{S}_k\mat{S}_k^{\tp}$ is extension matrix $\mat{X}_k$, \eqref{eq:MEMlikelihood} and \eqref{eq:RMMlikelihood} are equivalent with $\mat{C}^{\rvec{h}}=\frac{1}{z}\mat{I}_2$.
\end{Remark}

\subsection{Dynamic Model}
In general, there are no restrictions on the dynamic models for the temporal evolution of the state and shape parameters.  For the sake of simplicity, we here assume linear equations 
 according to 
\begin{eqnarray}
\rvec{r}_{k+1} &=& \mat{A}^{\rvec{r}}_{k}\rvec{r}_k+\rvec{w}^{\rvec{r}}_k\enspace,\\\label{eq:dyn_shape}
\rvec{p}_{k+1} &=& \mat{A}^{\rvec{p}}_{k}\rvec{p}_k+\rvec{w}^{\rvec{p}}_k\enspace,
\end{eqnarray}
where 
\begin{compactitem}
\item $\mat{A}^{\rvec{r}}_k$ and $\mat{A}^{\rvec{p}}_k$ are process matrices; 
\item $\rvec{w}^{\rvec{r}}_k$ and $\rvec{w}^{\rvec{p}}_k$ are zero-mean Gaussian process noises with covariance matrices $\mat{C}_{\rvec{r}}^{\rvec{w}}$ and  $\mat{C}_{\rvec{p}}^{\rvec{w}}$. 
\end{compactitem}

\section{Estimation}\label{sec:estimation}
This section presents closed-form expressions for the measurement and time update step based on the Kalman filter.
For this purpose, we factorize the joint density for  the object kinematics and extension similar to \cite{Feldmann2010,Lan2016b}, see also (\textbf{C2}) in the introduction.  
By this means, it is not necessary to maintain the cross-correlation between the kinematic state and the shape parameters.
However, it is important to note that interdependencies between the object kinematics and extension are  incorporated in the update and prediction formulas.

The derivation of a (nonlinear) Kalman filter   is particularly difficult due to the high nonlinearities and zero-mean multiplicative noise in the measurement equation. To solve these issues, we derive problem-tailored approximations of the required moments by means of combining linearization and analytic moment calculation techniques.

 \subsection{Measurement Update}
The measurements $\{\rvec{y}_k^{(j)}\}_{j=1}^{n_k}$   from time step $k$ are incorporated sequentially in the measurement update. 
For this purpose, let
 $$\hat{\rvec{r}}^{(i-1)}_{k} ,\;  \hat{\rvec{p}}^{(i-1)}_{k}  \text{ and }  \mat{C}^{\rvec{r}(i-1)}_{k}, \;  \mat{C}^{\rvec{p}(i-1)}_{k}.$$
denote the  estimates for the kinematic state $\hat{\rvec{r}}^{(i-1)}_{k}$ and shape parameters $\hat{\rvec{p}}^{(i-1)}_{k}$ plus the corresponding covariance matrices, having incorporated all measurements up to time $k-1$ plus the measurements $\{\rvec{y}_k^{(j)}\}_{j=1}^{i-1}$ from time $k$.
 
In the measurement update, the next measurement $\rvec{y}_k^{(i)}$ is incorporated in order to obtain the updated estimates
 $$\hat{\rvec{r}}_{k}^{(i)} ,\;  \hat{\rvec{p}}_{k}^{(i)}   \text{ and }  \mat{C}^{\rvec{r}(i)}_{k},  \;  \mat{C}^{\rvec{p}(i)}_{k}\enspace .$$

Note that -- according to this notation -- the predicted estimates for time $k$ are denoted as $(\bullet)^{(0)}_k$, correspondingly.

\begin{Remark}
It is important to note that the measurements from a single time scan are incorporated sequentially (in an arbitrary order). Due to approximations, slightly different results might be obtained for different orderings.
\end{Remark}

As  shown in \cite{Fusion12_Baum}, the object extent  cannot be estimated with a linear estimator that works with the actual measurement, 
i.e., the shape parameters do not change when updated with a single measurement $\rvec{y}_k^{(i)}$ in the Kalman filter framework.
For this reason, a pseudo-measurement is constructed  based on $\rvec{y}_k^{(i)}$ in order to update the shape parameters.
This can be seen as an uncorrelated transformation as discussed in \cite{Lan2015}.

\begin{table}
\caption{Measurement update of the MEM-EKF* algorithm. 
Source code: \protect\url{https://github.com/Fusion-Goettingen/}}
\label{tab:pseudo_code_meas}
\rule[0pt]{\columnwidth}{1pt}
\textbf{Input:}  Measurements $\{\rvec{y}_k^{(i)}\}_{i=1}^{n_k}$, predicted estimates  $\hat{\rvec{r}}^{(0)}_{k}$,  $\hat{\rvec{p}}^{(0)}_{k}$,  $\mat{C}^{\rvec{r}(0)}_{k}$, $\mat{C}^{\rvec{p}(0)}_{k}$, measurement noise covariance $\mat{C}^{\rvec{v}}$, $\mat{H}$ as defined in \eqref{eq:meas_kinematic}, multiplicative noise covariance $\mat{C}^{\rvec{h}}$, $\mat{F}$ and $\widetilde{\mat{F}}$ as defined in \eqref{eq:mat_F}

\textbf{Output:}  updated estimates $\hat{\rvec{r}}^{(n_k)}_{k} ,\;  \hat{\rvec{p}}^{(n_k)}_{k}  \text{ and }  \mat{C}^{\rvec{r}(n_k)}_{k}, \;  \mat{C}^{\rvec{p}(n_k)}_{k}$

\textbf{For} $i=1,\cdots, n_k$ \footnotesize
\begin{align*} 
\begin{array}{rcl}
\vect{\alpha&l_1&l_2}^{\tp}&=&\rvec{p}_k^{(i-1)}\\
\mat{S}=\vect{\mat{S_1}\\\mat{S_2}}&=&\vect{\cos{\alpha} & -\sin{\alpha}\\
\sin{\alpha} & \cos{\alpha}}\vect{l_1&0\\0&l_2}\\
\mat{J_1}&=& \vect{-l_1\sin{\alpha} &\cos{\alpha}&0\\-l_2\cos{\alpha}&0&-\sin{\alpha}}\\
\mat{J_2}&=&\vect{l_1\cos{\alpha}&\sin{\alpha}&0\\-l_2\sin{\alpha}&0&\cos{\alpha}}\\
 \mat{C}^{\text{\Romannum{1}}} &=& \mat{S} \mat{C}^{\rvec{h}}\mat{S}^{\tp}\\
\mat{C}^{\text{\Romannum{2}}}=[\epsilon_{mn}]  &=&  \trace{\mat{C}^{\rvec{p}^{(i-1)}}_k\mat{J}_n^{\tp}\mat{C}^{\rvec{h}}\mat{J}_m}  \text{ for }m,n=1,2\\
\mat{M}&=& \vect{2\mat{S_1}\mat{C}^{\rvec{h}}\mat{J_1}\\
					2\mat{S_2}\mat{C}^{\rvec{h}}\mat{J_2}\\
					\mat{S_1}\mat{C}^{\rvec{h}}\mat{J_2}+\mat{S_2}\mat{C}^{\rvec{h}}\mat{J_1}}\\
\bar{\rvec{y}}_k^{(i)}&=& \mat{H}\hat{\rvec{r}}_k^{i-1}\\
\mat{C}_k^{\rvec{r}\rvec{y}^{(i)}} &= &\mat{C}_k^{\rvec{r}(i-1)}\mat{H}^{\tp}\\
\mat{C}_k^{\rvec{y}^{(i)}} &=&
		\mat{H}\mat{C}_k^{\rvec{r}(i-1)}\mat{H}^{\tp} + \mat{C}^{\text{\Romannum{1}}}+\mat{C}^{\text{\Romannum{2}}}+\mat{C}^{\rvec{v}}\\
\rvec{Y}_k^{(i)} &=& \mat{F}\left((\rvec{y}^{(i)}_k-\bar{\rvec{y}}_k^{(i)})\otimes(\rvec{y}^{(i)}_k-\bar{\rvec{y}}_k^{(i)})\right)\\
\bar{\rvec{Y}}_k^{(i)}&=& \mat{F}\vecterize{\mat{C}_k^{\rvec{y}^{(i)}}}\\
\mat{C}_k^{\rvec{p}\rvec{Y}(i)} &=&\mat{C}_{k}^{\rvec{p}(i-1)}\mat{M}^{\tp}\\
\mat{C}_k^{\rvec{Y}(i)}&=&\mat{F}(\mat{C}^{\rvec{y}(i)}_k\otimes\mat{C}^{\rvec{y}(i)}_k)(\mat{F}+\widetilde{\mat{F}})^{\tp}\\
\hat{\rvec{r}}_k^{(i)} &=&\hat{\rvec{r}}_k^{(i-1)}+\mat{C}_k^{\rvec{r}\rvec{y}(i)}\left(\mat{C}_k^{\rvec{y}(i)}\right)^{-1}\left(\rvec{y}_k^{(i)}-\bar{\rvec{y}}_k^{(i)}\right)\\
\mat{C}_k^{\rvec{r}(i)}&=&\mat{C}_k^{\rvec{r}(i-1)}-\mat{C}_k^{\rvec{r}\rvec{y}(i)}\left(\mat{C}_k^{\rvec{y}(i)}\right)^{-1}\left(\mat{C}_k^{\rvec{r}\rvec{y}(i)}\right)^{\tp}\\
\hat{\rvec{p}}_k^{(i)}&=&\hat{\rvec{p}}_k^{(i-1)}+\mat{C}^{\rvec{p}\rvec{Y}(i)}_{k}\left(\mat{C}^{\rvec{Y}(i)}_{k}\right)^{-1}\left(\rvec{Y}_k^{(i)}-\bar{\rvec{Y}}^{(i)}_k\right)\\
\mat{C}^{\rvec{p}(i)}_k&=&\mat{C}^{\rvec{p}(i-1)}_{k}-\mat{C}^{\rvec{p}\rvec{Y}(i)}_{k}\left(\mat{C}^{\rvec{Y}(i)}_{k}\right)^{-1}\left(\mat{C}^{\rvec{p}\rvec{Y}(i)}_{k}\right)^{\tp}
\end{array}
\end{align*}
\textbf{End}

\rule[0pt]{\columnwidth}{1pt}
\end{table}

\subsubsection{Kinematic State Update}
The kinematic state estimate is updated according to  the  Kalman filter update equations using the actual measurement $\rvec{y}_k^{(i)}$, see  (\textbf{C3}),
\begin{eqnarray}
\bar{\rvec{y}}_k^{(i)}&=& \mat{H}\hat{\rvec{r}}_k^{(i-1)},\\
\hat{\rvec{r}}_k^{(i)}&=&\hat{\rvec{r}}_k^{(i-1)}+\mat{C}_k^{\rvec{r}\rvec{y}(i)}\left(\mat{C}_k^{\rvec{y}(i)}\right)^{-1}\left(\rvec{y}_k^{(i)}-\bar{\rvec{y}}_k^{(i)}\right),\label{eqn:kinupdate_mean}\\
\mat{C}_k^{\rvec{r}(i)}&=&\mat{C}_k^{\rvec{r}(i-1)}-\mat{C}_k^{\rvec{r}\rvec{y}(i)}\left(\mat{C}_k^{\rvec{y}(i)}\right)^{-1}\left(\mat{C}_k^{\rvec{r}\rvec{y}(i)}\right)^{\tp}\enspace.\label{eqn:kinupdate_cov}
\end{eqnarray}
The challenge is to find compact closed-form approximations to the required moments, i.e., the covariance of the measurement $\mat{C}_k^{\rvec{y}(i)}$ and   the cross-correlation $\mat{C}_k^{\rvec{r}\rvec{y}(i)}$ between the measurement and kinematic state.

The measurement equation \eqref{eq:meas_kinematic} is linear in the kinematic state but nonlinear in the shape parameters due to the shape matrix $\mat{S}_k$.
Linearizing $\mat{S}_k\rvec{h}_k^{(i)}$ with respect to $\rvec{p}_k$ at $\hat{\rvec{p}}_k^{(i-1)}$ and keeping $\rvec{h}_k^{(i)}$ as a random variable    (\textbf{C4}) gives us
\begin{equation}\label{eqn:sh}
\mat{S}_k\rvec{h}_k^{(i)} \approx \underbrace{\hat{\mat{S}}_k^{(i-1)}\rvec{h}_k^{(i)}}_{\text{\Romannum{1}}}+ \underbrace{\vect{\left(\rvec{h}_k^{(i)}\right)^{\tp} \widehat{\mat{J_1}}_{k}^{(i-1)}\\\left(\rvec{h}_k^{(i)}\right)^{\tp}\widehat{\mat{J_2}}_{k}^{(i-1)}}\left(\rvec{p}_k-\hat{\rvec{p}}_k^{(i-1)}\right)}_{\text{\Romannum{2}}}
\end{equation}
where $\hat{\bullet}_k^{(i-1)}$ denotes matrix $\bullet$ evaluated at the $(i-1)$-th shape estimate $\hat{\rvec{p}}_k^{(i-1)}$, $\mat{J_1}$ and $\mat{J_2}$ are the Jacobian matrices of the first row and second row of $\mat{S}$,  i.e.,
\begin{eqnarray}\label{eq:J1}
\mat{J_1}&=& \frac{\partial \mat{S_1}}{\partial \rvec{p}}=\vect{-l_{1}\sin{\alpha} &\cos{\alpha}&0\\-l_{2}\cos{\alpha}&0&-\sin{\alpha}},\\\label{eq:J2}
\mat{J_2}&=&\frac{\partial \mat{S_2}}{\partial \rvec{p}}=\vect{l_{1}\cos{\alpha}&\sin{\alpha}&0\\-l_{2}\sin{\alpha}&0&\cos{\alpha}},
\end{eqnarray}
with
\begin{eqnarray}
\mat{S_1}=\vect{l_1\cos{\alpha} & -l_2\sin{\alpha}}\text{ and } \mat{S_2}=\vect{l_1\sin{\alpha} & l_2\cos{\alpha}}.
\end{eqnarray}
Note that the terms \Romannum{1} and \Romannum{2} in \Eq{eqn:sh} are uncorrelated.
The covariance of $\mat{S}_k\rvec{h}_k^{(i)}$ is approximated as the sum of $\mat{C}^{\text{\Romannum{1}}}$ and $\mat{C}^{\text{\Romannum{2}}}$, where
\begin{eqnarray}
\mat{C}^{\text{\Romannum{1}}} & = & \hat{\mat{S}}_k^{(i-1)}\mat{C}^{\rvec{h}}(\hat{\mat{S}}_k^{(i-1)})^{\tp}\enspace,\\\label{eq:cov_2}
\underbrace{[\epsilon_{mn}]}_{\mat{C}^{\text{\Romannum{2}}}} & = & \trace{\mat{C}^{\rvec{p}^{(i-1)}}_k\left(\widehat{\mat{J}_n}^{(i-1)}_k\right)^{\tp}\mat{C}^{\rvec{h}}
\widehat{\mat{J}_m}^{(i-1)}_k}\enspace,
\end{eqnarray}
for $m,n \in \{1,2\}$. 

The derivation of \eqref{eq:cov_2} is shown in Appendix \ref{app:cov_2}. 
The cross-covariance and covariance are
\begin{eqnarray}
\mat{C}_k^{\rvec{r}\rvec{y}^{(i)}} &=& \mat{C}_k^{\rvec{r}(i-1)}\mat{H}^{\tp}\\\label{eq:meas_cov}
\mat{C}_k^{\rvec{y}^{(i)}} &=&
		\mat{H}\mat{C}^{\rvec{r}(i-1)}\mat{H}^{\tp} + \mat{C}^{\text{\Romannum{1}}}+\mat{C}^{\text{\Romannum{2}}}+\mat{C}^{\rvec{v}} \enspace .
\end{eqnarray}

\subsubsection{Shape  Update} 
A pseudo-measurement is constructed using the 2-fold Kronecker product (\textbf{C3}).
For a two-dimensional  vector $\rvec{y}=\vect{y_1&y_2}^{\tp}$, its 2-fold Kronecker product $\otimes$ is defined as
\begin{equation}\label{eq:2foldKronecker}
\rvec{y}\otimes\rvec{y}=\vect{y_1^2&y_1y_2&y_2y_1&y_2^2}^\tp \enspace .
\end{equation}
Furthermore, each measurement is shifted by the expected measurement, and multiplied by a matrix 
\begin{equation}\label{eq:mat_F}
\mat{F}=\vect{1&0&0&0\\0&0&0&1\\0&1&0&0&}\text{ or }\widetilde{\mat{F}}=\vect{1&0&0&0\\0&0&0&1\\0&0&1&0&}
\end{equation}
 to remove the duplicate element resulting from the 2-fold Kronecker product.
All told, the pseudo-measurement is
\begin{equation}\label{eq:pseudo_meas}
\rvec{Y}_k^{(i)} = \mat{F}\left((\rvec{y}^{(i)}_k-\bar{\rvec{y}}_k^{(i)})\otimes(\rvec{y}^{(i)}_k-\bar{\rvec{y}}_k^{(i)})\right)\enspace .
\end{equation}
Note that \eqref{eq:pseudo_meas} is an uncorrelated conversion (c.f., Theorem 3 in \cite{Lan2015}), which means the pseudo-measurement is uncorrelated with the actual measurement.

The shape parameters are updated with the pseudo-measurement $\rvec{Y}_k^{(i)}$  using the Kalman filter update formulas 
\begin{eqnarray}
		\rvec{p}_k^{(i)}&=&\hat{\rvec{p}}_k^{(i-1)}+\mat{C}^{\rvec{p}\rvec{Y}(i)}_{k}\left(\mat{C}^{\rvec{Y}(i)}_{k}\right)^{-1}\left(\rvec{Y}_k^{(i)}-\bar{\rvec{Y}}^{(i)}_k\right)\qquad\label{eqn:shapeupdate_mean}\\
		\mat{C}^{\rvec{p}(i)}_k&=&\mat{C}^{\rvec{p}(i-1)}_{k}-\mat{C}^{\rvec{p}\rvec{Y}(i)}_{k}\left(\mat{C}^{\rvec{Y}(i)}_{k}\right)^{-1}\left(\mat{C}^{\rvec{p}\rvec{Y}(i)}_{k}\right)^{\tp}\enspace\label{eqn:shapeupdate_cov}
\end{eqnarray}
where $\bar{\rvec{Y}}^{(i)}$ denotes the predicted pseudo-measurement, $\mat{C}^{\rvec{Y}(i)}_{k}$ is the covariance of the pseudo-measurement, and $\mat{C}^{\rvec{p}\rvec{Y}(i)}_{k}$ is the cross-covariance between the pseudo-measurement and the shape parameters.

By constructing the pseudo-measurement in this way, the expected pseudo-measurement happens to consist of all centralized second moments of the actual measurements, which can be extracted directly from \eqref{eq:meas_cov}, see  (\textbf{C5}).
To show this, we introduce the $\text{vect}$-operator, which  constructs a column vector from a matrix by stacking its column vectors. 
Given the covariance matrix of measurement 
$\mat{C}^{\rvec{y}(i)}_k =\vect{c_{11}&c_{12}\\c_{12}&c_{22}},$ 
a $\text{vect}$-operator gives us
\begin{equation}
\vecterize{ \mat{C}^{\rvec{y}(i)}_k} =\vect{c_{11} & c_{12}&c_{12}&c_{22}}^{\tp} \enspace ,
\end{equation}
which equals
\begin{equation}
\expect{(\rvec{y}^{(i)}_k-\bar{\rvec{y}}_k^{(i)})\otimes(\rvec{y}^{(i)}_k-\bar{\rvec{y}}_k^{(i)})}\enspace .
\end{equation}
The expected $i$-th pseudo-measurement is
\begin{equation}
\bar{\rvec{Y}}_k^{(i)}= \mat{F}\vecterize{\mat{C}^{\rvec{y}(i)}_k}\enspace.
\end{equation}
The predicted pseudo-measurement covariance is
\begin{eqnarray}
	\label{eq:pseudo_cov}
		\mat{C}_k^{\rvec{Y}(i)}&=&\vect{2c_{11}^2 &2c_{12}^2 & 2c_{11}c_{12}\\
									2c_{12}^2&2c_{22}^2 & 2c_{22}c_{12}\\
									2c_{11}c_{12} & 2c_{22}c_{12} &c_{11}c_{22}+c_{12}^2 },\\\label{eq:pseudo_cov_simp}
								&=& \mat{F}(\mat{C}^{\rvec{y}(i)}_k\otimes\mat{C}^{\rvec{y}(i)}_k)(\mat{F}+\widetilde{\mat{F}})^{\tp} \enspace .
\end{eqnarray}
Equation \eqref{eq:pseudo_cov} is obtained using Isserlis's theorem \cite{Isserlis1918} (see Appendix  \ref{app:pseudo-cov}).
Equation \eqref{eq:pseudo_cov_simp} is a compact formulation of \eqref{eq:pseudo_cov}.

The cross-covariance between the pseudo-measurement and the shape parameters is approximated by linearization of \eqref{eq:pseudo_meas} according to
\begin{equation}
\label{eq:pseudo_cross_cov}
\mat{C}_k^{\rvec{p}\rvec{Y}(i)} =\mat{C}_{k}^{\rvec{p}(i-1)}\left(\hat{\mat{M}}_k^{(i-1)}\right)^{\tp}\enspace,
\end{equation}
with 
\begin{equation}\label{eq:matrixM}
\mat{M}= 
			\vect{2\mat{S_1}\mat{C}^{\rvec{h}}\mat{J_1}\\
					2\mat{S_2}\mat{C}^{\rvec{h}}\mat{J_2}\\
					\mat{S_1}\mat{C}^{\rvec{h}}\mat{J_2}+\mat{S_2}\mat{C}^{\rvec{h}}\mat{J_1}}\enspace.
\end{equation}
The derivation of \eqref{eq:matrixM} is shown in the Appendix \ref{app:jacobian}. 
The pseudo code of measurement update is given in Table \ref{tab:pseudo_code_meas}.

\subsection{Time Update}
As the temporal evolution of both the kinematic state and the  shape parameters follow a linear model, the time update can be performed with  the standard Kalman filter time update formulas, i.e.,
\begin{eqnarray}
\hat{\rvec{r}}_{k+1}^{(0)}&=& \mat{A}^{\rvec{r}}_k\hat{\rvec{r}}_k^{(n_k)},\\
\mat{C}_{k+1}^{\rvec{r}(0)} &=& \mat{A}^{\rvec{r}}_{k}\mat{C}_{k}^{\rvec{r}(n_k)}(\mat{A}^{\rvec{r}}_{k})^{\tp}+\mat{C}_{\rvec{r}}^{\rvec{w}}.
\end{eqnarray}  
and
\begin{eqnarray}
\hat{\rvec{p}}_{k+1}^{(0)}&=& \mat{A}^{\rvec{p}}_k\hat{\rvec{p}}_k^{(n_k)}\enspace,\\
\mat{C}_{k+1}^{\rvec{p}(0)} &=& \mat{A}^{\rvec{p}}_{k}\mat{C}_{k}^{\rvec{p}(n_k)}(\mat{A}^{\rvec{p}}_{k})^{\tp}+\mat{C}_{\rvec{p}}^{\rvec{w}}\enspace.
\end{eqnarray}

\section{Evaluation}\label{sec:eval}

\definecolor{cyan}{rgb}{0.00000,1.00000,1.00000}
\definecolor{magenta}{rgb}{1.00000,0.00000,1.00000}
\begin{figure*}
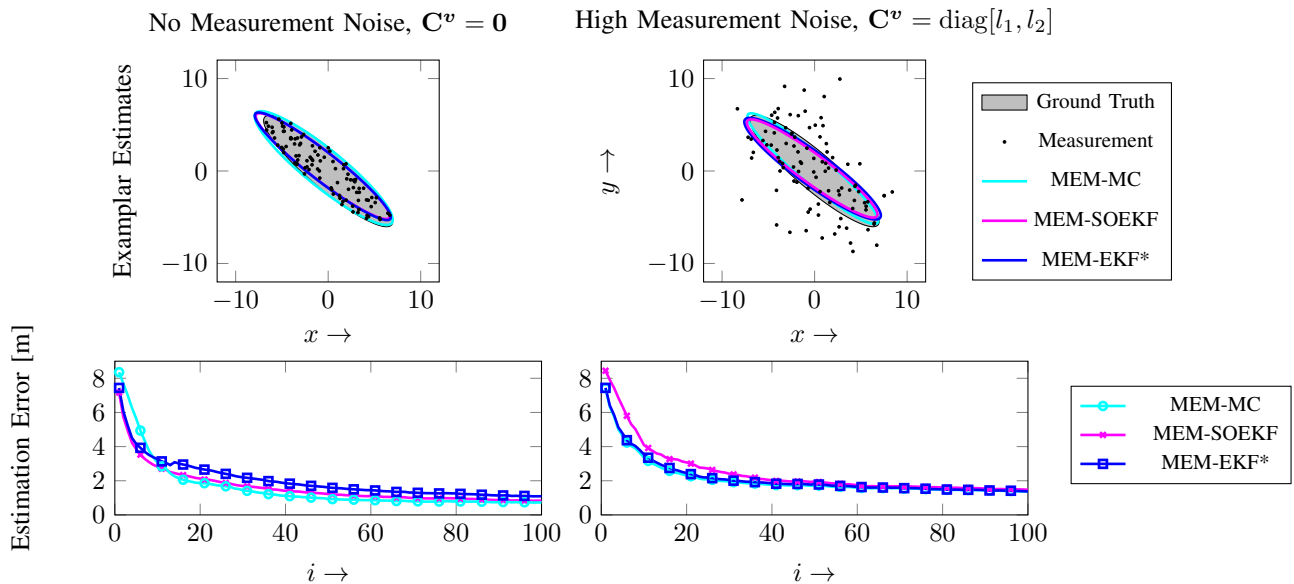

\centering
	\begin{tikzpicture}
	\tikzset{mark size=.5}
		\begin{groupplot}[group style = {group size = 2 by 2, vertical sep=30pt, horizontal sep = 100pt}, width = 0.25\textwidth, height = 0.25\textwidth]
			\nextgroupplot[ title = {No Measurement Noise, $\mat{C}^{\rvec{v}}=\mat{0}$},ylabel={Examplar  Estimates},xlabel={$x\rightarrow$},
			legend style = {font=\footnotesize,at={(3.4,.9)}, anchor=north west,row sep = 4pt, legend columns = 1},ymin=-12,ymax=12,xmin=-12,xmax=12]
				\input{measCov0_leg.tex}
				\label{fig:static_meas_meas_noise_0}
			\nextgroupplot[title = {High Measurement Noise, $\mat{C}^{\rvec{v}}=\diag[l_1,l_2]$},xlabel={$x\rightarrow$},ylabel={$y\rightarrow$},
							ymin=-12,ymax=12,xmin=-12,xmax=12]
				\input{measCov1.tex}
				\label{fig:static_meas_meas_noise_2}
			\nextgroupplot[ylabel={Estimation Error [m]},xlabel={$i\rightarrow$},ymin=0,ymax=9,xmin=0,xmax=100,width=0.4\textwidth,height=0.2\textwidth]
				\input{rmse_measCov0.tex}
				\label{fig:static_state_meas_noise_0}
			\nextgroupplot[xlabel={$i\rightarrow$},
			legend style = {font=\footnotesize,at={(1.1,.2)}, anchor=south west, column sep = 8pt, legend columns = 1}, ymin=0,ymax=9,xmin=0,xmax=100,,width=0.4\textwidth,height=0.2\textwidth]
				\input{rmse_measCov1.tex}
				\label{fig:static_state_meas_noise_1}
		\end{groupplot}
	\end{tikzpicture}
	\caption{Simulation with a stationary ellipse. The first row shows the ground truth, exemplar measurements, and the estimates after $100$ measurement updates. The bottom row plots the root mean squared  Gaussian Wasserstein distance averaged over $100$ runs. 
	}
		\label{fig:static_ellipse_result}
\end{figure*}
\begin{figure*}
\centering
\includegraphics[width=.88\textwidth]{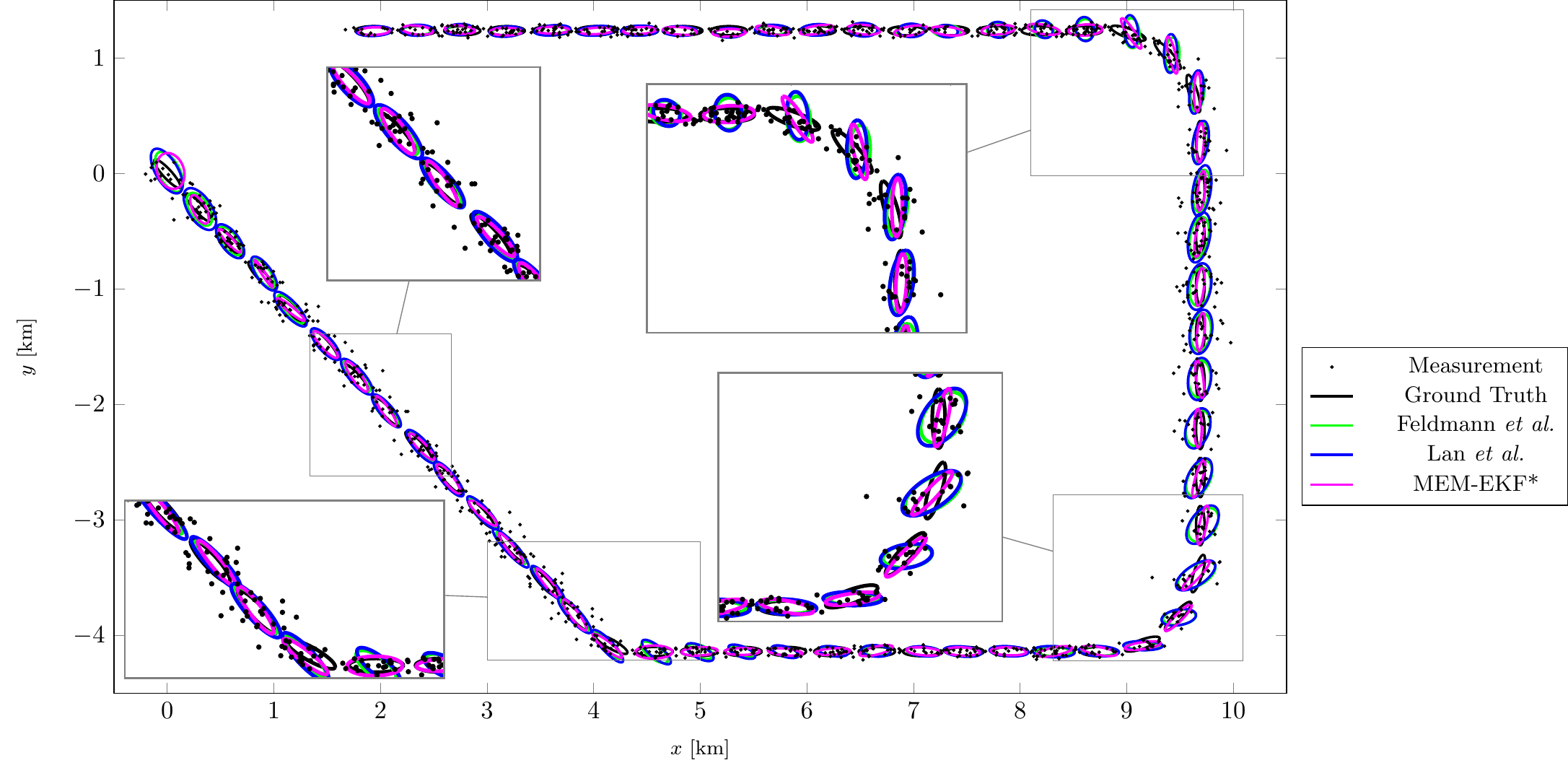}
\caption{The measurements, trajectory, and estimation results of a single example run.}
\label{fig:snippet}
\end{figure*}
\begin{figure}
    \centering
	\begin{tikzpicture}
	\tikzset{mark size=0.5}
		\begin{axis}[ylabel={ \footnotesize[m]},xlabel={\footnotesize $k\rightarrow$},width=.4\textwidth,height=1.3in,scale only axis,xmin=0,xmax=180,xmajorgrids,ymin=0,ymax=160,ymajorgrids,axis background/.style={fill=white},
		legend style = {at={(1,1)},font=\footnotesize, column sep = 2pt, legend columns = 3, }]
		\input{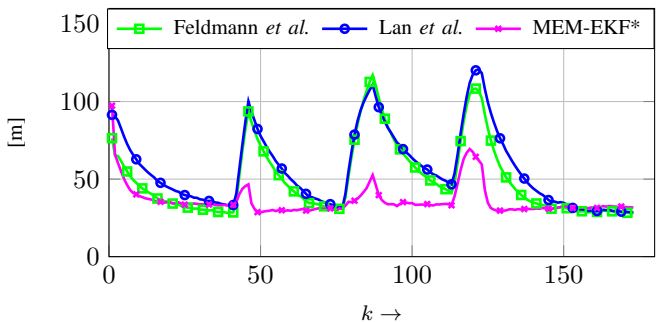}
		\end{axis}
	\end{tikzpicture}
    \caption{Extent error based on the mean squared Gaussian Wasserstein distance.}
    \label{fig:rmse_extent}
\end{figure}

In this section, 
 we first evaluate the accuracy of the developed moment approximations. 
 Then, the benefits of the developed shape tracker are demonstrated  with respect to the random matrix approach in a simple extension dynamics.
 In the end, we integrate turn rate estimation and tested in a scenario in which object extent is coupled with its kinematics.

We assess location and extent errors simultaneously with a single score by means of the  Gaussian Wasserstein distance \cite{Givens1984} as proposed  \cite{MFI16_Yang}.
It is very important to note that orientation and axes-lengths errors are combined in a single scalar value.

The Gaussian Wasserstein distance compares two ellipses according to 
\begin{multline}\label{eq:gaussian_wasserstein_def}
			d(\rvec{\mu}_1, \mat{\Sigma}_1, \rvec{\mu}_2, {\mat{\Sigma}_2})^2 = \parallel \rvec{\mu}_1 - \rvec{\mu}_{2}\parallel^{2} \\+
			 \trace{\mat{\Sigma}_1+\mat{\Sigma}_{2} -
		 		2\sqrt{\sqrt{\mat{\Sigma}_1}\mat{\Sigma}_{2}\sqrt{\mat{\Sigma}_1}}} \enspace ,
		\end{multline}		
 where the ellipses are specified by their locations  $\rvec{\mu}_1\in \IR^2$ and $\rvec{\mu}_2 \in \IR^2 $
and SPD shape matrices  $\mat{\Sigma}_1 \in \IR^{2\times2}$ and $\mat{\Sigma}_2 \in \IR^{2\times2}$.
In this case, the first ellipse is the ground truth and the second one is the extended object tracking method estimate.
Note that an SPD shape matrix is computed using $\hat{\mat{S}}_k\hat{\mat{S}}_k^{\tp}$.

\subsection{Evaluation of Moment Approximations}
First, we evaluate the quality of the proposed  moment approximations for the kinematic state \Eq{eqn:kinupdate_mean} and \Eq{eqn:kinupdate_cov}, and the shape parameters \Eq{eqn:shapeupdate_mean} and \Eq{eqn:shapeupdate_cov} compared to the  Monte Carlo moment approximation  and our Second-Order EKF \cite{Yang2016}, which requires the calculation of Hessian matrices. Both methods are computationally much more complex than the proposed tracker.
As we focus on the moment approximations of the measurement update, we restrict ourselves to a scenario with a non-moving object.
The considered object is located in the origin with semi-axes lengths $2$ and $9$ meters and it is counter-clockwise rotated  $\frac{\pi}{3}$. 
The prior for the shape parameters is 
\begin{align}\notag
\hat{\rvec{r}}_1^{(0)}&=\vect{1&1},  &\mat{C}_1^{{r}(0)}&=\diag{\vect{1 &1}},\\\notag
\hat{\rvec{p}}_1^{(0)}&=\vect{0&2&12},&\mat{C}_1^{{p}(0)}&=\diag{\vect{1&4 &9}}
\end{align}
for all three methods. 
Two different measurement noise covariance matrices  are evaluated and the simulation results are shown in Fig. \ref{fig:static_ellipse_result}.

As expected, the Monte Carlo moments approximation outperforms the analytic approaches in both low and high measurement noises scenarios.
In case of high measurement noise, our moments approximation is almost as good as the Monte Carlo approximation unexpected.
 \Fig{fig:static_ellipse_result} shows that  there are no significant visual differences between the methods.
We can conclude that the derived moment approximations  nearly matches the true exact moments in low and high noise scenarios.

\subsection{Comparison with Random Matrix Approaches using Constant Velocity Motion Models} \label{sec:sim1}

In the second simulation, we compare our algorithm with the two random matrix approaches  by Feldmann \etal \cite{Feldmann2010} and Lan \etal \cite{Lan2016b}.

The considered scenario involves a target object with an unknown extent. However, only its orientation is changing over time, its semi-axes are fixed.
With the MEM-EKF* we can assume  a low system noise on the semi-axes lengths, and high system noise on the orientation. However, the random matrix approaches \cite{Feldmann2010,Lan2016b} cannot model this scenario precisely (as there is only a single parameter for the extent uncertainty). 
For this reason, the MEM-EKF* is able to outperform the random matrix approaches in this scenario. 
The true track is similar as in \cite{Feldmann2010} and \cite{Lan2016b}. 
The extended object has diameters of 340$m$ and 80$m$.  It starts at the  coordinate origin and  it moves with a constant speed of 50$km/h$. 
At each time step, measurement sources are generated from a uniform distribution on the elliptical extent. 
The number of measurements per time scan is drawn from a Poisson distribution with mean $20$ as in \cite{Lan2016b}. 
The variances of the measurement noise are 10000$m^2$ and 400$m^2$ for each dimension.

We set
$\tau$ to $50$ in the approach of Feldmann \etal; $\delta$ to $40$ in the approach of Lan \etal; $v$ to $56$ in both random matrix approaches.  
The extension transition matrix in Lan \etal's approach is $\frac{1}{\delta_k}\mat{I}_2$.
For our method,  the prior of the shape variables is specified by the covariance matrix
$\mat{C}^{\rvec{p}}_{0}=\diag{[1,70^2,70^2]}$.  
The process noise covariance is set to
$\mat{C}^{\rvec{w}}_{\rvec{p}}=\diag{[0.1,1,1]}$, and 
the transition matrix is
$\mat{A}_k^{\rvec{p}}=\mat{I}_3$.
The process noise covariance for the kinematic state is  $\diag{[100,100,1,1]}$ for all three estimators.

Measurements, trajectory, and estimation results of an example run   are depicted in Fig.~\ref{fig:snippet}. 
Both random matrix trackers have worse results during turns. 
The extent error according to the Gaussian Wasserstein distance is depicted in Fig.~\ref{fig:rmse_extent}. 
From Fig.~\ref{fig:rmse_extent} we can  conclude that Feldmann \etal and Lan \etal perform similarly overall.
This is expected as both random matrix methods are the same if no special dynamic model for the extend is used. 
For all three turns the proposed tracker has a lower error compared to both random matrix approaches due to the aforementioned assumption of  (nearly) static semi-axis.

\subsection{Comparison with Random Matrix Approaches using Constant Turn Motion Models}

  \begin{figure}
     \centering
     \input{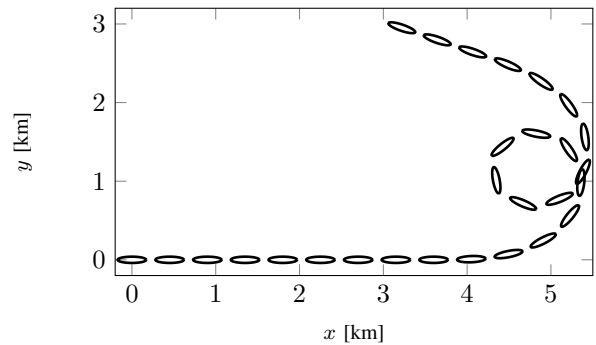}
     \caption{True trajectory of variable turn simulation.}
     \label{fig:GT_turn}
 \end{figure}
 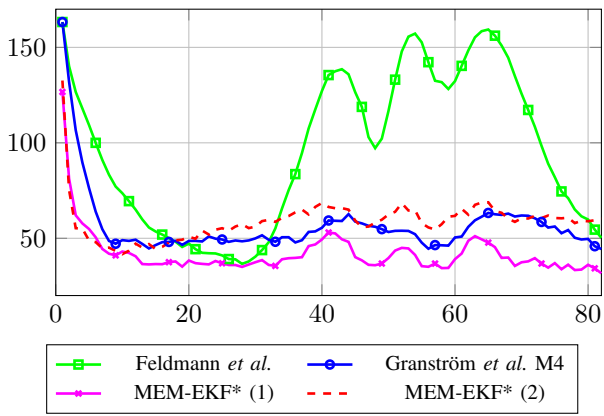
\begin{figure}
     \centering
%
%
\definecolor{mycolor1}{rgb}{1.00000,0.00000,1.00000}%
\begin{tikzpicture}

\begin{axis}[%
width=0.4\textwidth,
height=1.5in,
at={(1.011in,0.642in)},
grid = major,
scale only axis,
xmin=0,
xmax=82,
ymin=20,
ymax=170,
axis background/.style={fill=white},
title style={font=\bfseries},
legend style = {font=\footnotesize, column sep = 10pt, legend columns = 2, at={(0.47,-0.17)},anchor=north}
]

\addplot [color=green,,mark=square,mark repeat=5,mark phase=1,mark size=1.5pt,line width=1.0pt]
  table[row sep=crcr]{%
1	163.277555090833\\
2	140.141502371904\\
3	126.654907479932\\
4	117.999574473311\\
5	109.307869705009\\
6	99.9996465237147\\
7	91.001637272582\\
8	83.2893128968516\\
9	77.1007239221099\\
10	73.6218683191009\\
11	69.4611745710684\\
12	65.0140687959665\\
13	59.9088779428971\\
14	55.653782441681\\
15	53.5682869686926\\
16	51.9434635329197\\
17	50.079120317732\\
18	48.2047201072953\\
19	45.4709318131674\\
20	47.2130288563301\\
21	44.2970277418854\\
22	42.3901215923484\\
23	42.16713341907\\
24	42.0531816581018\\
25	41.0340050725246\\
26	39.1211666413689\\
27	38.9794715565596\\
28	36.5774561651869\\
29	37.8754292004813\\
30	40.3824857410279\\
31	43.8160256845703\\
32	47.9108145544363\\
33	55.1649139184048\\
34	65.5975411226897\\
35	75.2230675445807\\
36	83.6115023987972\\
37	94.6527991805757\\
38	107.484295028686\\
39	117.470154829135\\
40	127.144071717352\\
41	135.491414741611\\
42	137.490630059281\\
43	138.565256431293\\
44	135.908527020825\\
45	126.535859965503\\
46	118.831790378138\\
47	103.017842507898\\
48	97.2053170411777\\
49	102.196142778561\\
50	116.662299480744\\
51	133.087108152439\\
52	147.861422813083\\
53	155.613941824787\\
54	157.216793267913\\
55	152.639473003513\\
56	142.245451597052\\
57	132.509618764563\\
58	131.779933129847\\
59	128.203115722196\\
60	132.313299584847\\
61	140.320500928789\\
62	148.398222154181\\
63	155.263832691227\\
64	158.471501626132\\
65	159.358039898477\\
66	156.132410897697\\
67	150.967236572045\\
68	144.577054914818\\
69	134.952345538427\\
70	125.742589903419\\
71	117.216631381312\\
72	108.425830422937\\
73	98.4497468194119\\
74	89.0801011578899\\
75	82.0143541702187\\
76	74.549832732324\\
77	71.0248395619599\\
78	65.0901448891678\\
79	61.5226857375872\\
80	59.8237740884741\\
81	54.5313969798119\\
82	50.2032619827416\\
};
\addlegendentry{Feldmann \etal}

\addplot [color=blue, solid,mark=o,mark repeat=8,mark phase=1,mark size=1,solid,mark size=1.5pt,line width=1.0pt]
  table[row sep=crcr]{%
1	163.277555090833\\
2	132.00433035178\\
3	106.608848490161\\
4	89.715127666277\\
5	76.285839176502\\
6	63.9681802006367\\
7	54.753958786716\\
8	48.6046729392228\\
9	47.1397889224836\\
10	49.019672455073\\
11	48.6368111735882\\
12	49.2561774733438\\
13	47.0513230578575\\
14	44.6703206413789\\
15	46.4274768876083\\
16	47.967615283705\\
17	48.0947121137254\\
18	48.6918616154781\\
19	46.5480594919068\\
20	48.6092640903834\\
21	48.6363766645586\\
22	48.3376308004361\\
23	50.9647179052448\\
24	49.7944142284694\\
25	49.3679071723733\\
26	48.1333455672138\\
27	48.8503814298896\\
28	48.4215126617001\\
29	48.8746508134793\\
30	49.9386323727096\\
31	51.567858385319\\
32	49.2233558252482\\
33	48.1706200806076\\
34	50.5459532351172\\
35	50.5599786383087\\
36	47.8208681388342\\
37	48.9856426955877\\
38	53.3311827799184\\
39	53.6024951880429\\
40	55.5010032256752\\
41	59.3180697660612\\
42	59.1750853095004\\
43	59.0378349311862\\
44	62.6369245114027\\
45	58.7264246618694\\
46	57.8089397719163\\
47	56.5314462081745\\
48	56.0350936537384\\
49	54.7845888445106\\
50	53.2908172332692\\
51	53.8885608088773\\
52	53.9484365859387\\
53	53.8477529283883\\
54	52.1432350023109\\
55	47.7162196445465\\
56	44.5262485396074\\
57	46.4344431314955\\
58	46.3295046962934\\
59	46.1553529105493\\
60	50.4319220777759\\
61	50.8926821637664\\
62	56.0071738497139\\
63	59.4105195257739\\
64	61.2837978907012\\
65	63.2730583367012\\
66	62.5779337803111\\
67	62.2451331456153\\
68	62.9304628083765\\
69	61.3968023942382\\
70	61.9375817243581\\
71	61.8010824858717\\
72	60.4784795489295\\
73	58.5536533340811\\
74	55.6086973753642\\
75	56.1660469260643\\
76	52.6378274212019\\
77	54.3182259876048\\
78	50.2639095274815\\
79	49.3621283514537\\
80	49.6836129943343\\
81	45.8420005256413\\
82	44.3988825532911\\
};
\addlegendentry{ Granstr\"om \etal M4}
\addplot [color=mycolor1,mark=x,mark repeat=8,mark phase=1,mark size=1,solid,mark size=1.5pt,line width=1.0pt]
  table[row sep=crcr]{%
1	126.645401189589\\
2	81.2305120172995\\
3	62.0241136465077\\
4	57.7826085750708\\
5	54.8735948959759\\
6	50.5140047908663\\
7	44.4025485247005\\
8	41.9155409001108\\
9	41.0548985169273\\
10	43.0626892117394\\
11	42.3324258793162\\
12	40.5771344340682\\
13	36.4927215134422\\
14	36.3160357541712\\
15	36.4938845764005\\
16	36.4259253221199\\
17	37.4905001647285\\
18	37.8655034834502\\
19	35.141202571619\\
20	38.3515398148221\\
21	36.9669555215911\\
22	36.5917723236514\\
23	36.3228677142508\\
24	38.0976262234517\\
25	36.9100933696587\\
26	35.995294086434\\
27	35.9981981573659\\
28	34.999466580432\\
29	36.2534271017398\\
30	37.4565576633384\\
31	38.5812263836869\\
32	36.035314086887\\
33	35.5352711788595\\
34	39.0085921030181\\
35	39.5607943654969\\
36	39.7352119158367\\
37	40.0877140626681\\
38	45.9234465647026\\
39	46.7566227222394\\
40	49.3807645582469\\
41	53.0269766090827\\
42	52.5486553589802\\
43	49.4555682750057\\
44	48.1271305000598\\
45	40.1376057187171\\
46	38.4493265714267\\
47	36.0928421823238\\
48	35.8854269637761\\
49	36.7749116692162\\
50	39.6493695464615\\
51	43.5553538207785\\
52	44.9807849800012\\
53	44.6785534616558\\
54	41.2748680564863\\
55	35.3990791604092\\
56	35.0645033224452\\
57	36.7496659107232\\
58	34.3728876239134\\
59	34.4597435399717\\
60	39.0691985780207\\
61	42.0112645270397\\
62	48.7030955788251\\
63	51.124113138836\\
64	49.4489050633086\\
65	47.7396886696568\\
66	45.3505507621403\\
67	39.7344827041515\\
68	39.91262661994\\
69	35.5125821774832\\
70	37.4272400224341\\
71	37.9063782405921\\
72	39.5982519037469\\
73	36.7780899387598\\
74	34.6215826527319\\
75	36.9799571717114\\
76	33.7734612147919\\
77	36.1735767129435\\
78	33.3536535757449\\
79	33.6133798807216\\
80	36.1196818793738\\
81	34.2477371625849\\
82	31.6733804671156\\
};
\addlegendentry{MEM-EKF* (1)}

\addplot [color=red,dashed,line width=1.0pt]
  table[row sep=crcr]{%
1	132.62423982344\\
2	75.3136909012877\\
3	55.2788248178055\\
4	55.0519633865403\\
5	48.8512775197228\\
6	47.9934717239293\\
7	44.7498625377594\\
8	45.1055610789066\\
9	43.4118944050135\\
10	41.6673680864887\\
11	43.5075287239435\\
12	47.1621675511951\\
13	45.1326899714655\\
14	46.9491525055293\\
15	44.3621310178739\\
16	45.4935701075775\\
17	47.7138734234611\\
18	48.9323020394512\\
19	49.2469110531373\\
20	50.8108312145704\\
21	49.7385764008194\\
22	54.0345263546368\\
23	51.9606772828391\\
24	54.4893859440777\\
25	55.2733421483107\\
26	53.9414870946784\\
27	56.1555682660833\\
28	57.4006825034566\\
29	55.3101001131275\\
30	56.1171059872354\\
31	58.994253019613\\
32	59.6359894020866\\
33	58.6029318052782\\
34	60.6473203626034\\
35	61.6480160512122\\
36	63.5793839264146\\
37	65.0543041117332\\
38	62.4938902412018\\
39	66.4310384769742\\
40	68.4852036270877\\
41	66.4062988741914\\
42	65.7399599246346\\
43	65.7907087538977\\
44	65.0103466680264\\
45	59.320035895438\\
46	55.7901150831153\\
47	56.1437770345613\\
48	58.2074268692737\\
49	59.7185455060129\\
50	62.4668981890745\\
51	63.8535174070364\\
52	68.378438472078\\
53	64.5557415174053\\
54	64.6828908121093\\
55	57.9276094756989\\
56	55.6786105423798\\
57	54.5824728393727\\
58	57.3892593227836\\
59	61.3053146372479\\
60	61.7225622038532\\
61	65.4470143093257\\
62	62.3846728191875\\
63	67.5282856969208\\
64	68.7237034896003\\
65	68.853733357962\\
66	64.285300441321\\
67	63.3511227669979\\
68	60.2226877919826\\
69	58.5986604558712\\
70	60.4433311371682\\
71	60.3671551577611\\
72	61.6841941498267\\
73	60.5744778020664\\
74	61.9146883202301\\
75	60.6229803173172\\
76	60.4624513875491\\
77	60.7648077482981\\
78	57.9448081265741\\
79	59.161412751812\\
80	58.9204385715894\\
81	59.4507364729741\\
82	60.340088626202\\
};
\addlegendentry{MEM-EKF* (2)}
\end{axis}

\end{tikzpicture}%
     \caption{Mean squared Gaussian Wasserstein distance for 100 runs.}
     \label{fig:turn_err}
 \end{figure}
For a  manoeuvring target, the object orientation is typically coupled with the turn rate \cite{Granstroem2013a}. 
Assuming that the size of the extended object is constant (plus noise), we can integrate the correlation between orientation and turn rate in the prediction of the MEM-EKF* according to
\begin{eqnarray}
\hat{\rvec{p}}_{k+1}^{(0)}&=& \mat{B}\hat{\rvec{r}}^{(n_k)}_{k} + \hat{\rvec{p}}_k^{(n_k)}\enspace,\\
\mat{C}_{k+1}^{\rvec{p}(0)} &=&\mat{B}\mat{C}_{k}^{\rvec{r}(n_k)}\mat{B}^{\tp}+ \mat{C}_{k}^{\rvec{p}(n_k)}+\mat{C}_{\rvec{p}}^{\rvec{w}},
\end{eqnarray} 
where $\mat{B}= \vect{\mat{0}_{1\times4}&T\\\mat{0}_{2\times4} &\mat{0}_{2\times1}}$ picks out turn rate $\hat{\Omega}_k^{(n_k)}$.

In this simulation, a target object that follows a variable turn-rate model is simulated, similar to \cite{Granstroem2013a}. 
The diameters of the simulated object are $170m$ and $40m$ and  the true track is shown in Fig.~\ref{fig:GT_turn}. In the first 25 time steps, the object moves with a constant velocity of $150m/s$. Afterwards, its turn rate increases from 0 to 20 degree per second in 20 time steps, and decreases to 0 in 20 time steps. 
The object evolves additional   5 time steps according to a  constant velocity model in the end of the trajectory. The number of measurements for each time step is Poisson distributed with mean $20$ and the measurement noise covariance is
$\mat{C}^v = \diag[10000\quad400]$. 
We compare the MEM-EKF*  with the random matrix approach M4 proposed \cite{Granstroem2013a}.
M4 approximates the density of the  object extent  with  an inverse Wishart density by minimizing Kullback-Leibler divergence in the prediction step.  
As a baseline, we also include the random matrix approach from \cite{Feldmann2010}, which does not incorporate the turn-rate.

For all estimators, the object kinematics is modeled as a constant turn model  and the prior  is
\begin{eqnarray}
\hat{\rvec{r}}_{1}^{(0)}&=&\vect{100&100& 100&20&0.001}^{\tp}\\ \mat{C}_1^{\rvec{r}(0)}&=&\diag[1600\mat{I}_2\quad16\mat{I}_2\quad0.001].
\end{eqnarray}
The parameters for the random matrix approaches are $v=56$, $ \tau = 5$, and $T = 1s$. 
The prior for our shape variables is
\begin{equation}
\hat{\rvec{p}}_{1}^{(0)}=\vect{\frac{\pi}{3}&200&90}^{\tp},\quad
\mat{C}_1^{\rvec{p}(0)}=\diag[0.2\enspace360\mat{I}_2].
\end{equation}
The process noise covariace matrices for the location, velocity, and turn rate are
$$\mat{C}_{\rvec{r}}^{\rvec{w}}=\diag[1000\mat{I}_2\quad100\mat{I}_2],\quad \mat{C}_{\Omega}^{\rvec{w}}=0.001.\quad$$

The MEM-EKF* constrains the temporal evolution of the extent  via process a  suitable noise covariance for the shape parameters. 
In the same way as the simulation in \Sec{sec:sim1}, the process noise covariance matrix is tuned such that the semi-axes are only allowed to change slightly over time.

To  demonstrate this effect, we choose two different process noise covariance matrices for our tracker and refer them as
\begin{eqnarray}
\text{MEM-EKF* (1)}&\text{ with }&\mat{C}_{\rvec{p}}^{\rvec{w}}=\diag[.01 \quad \mat{I}_2]\\
\text{MEM-EKF* (2)}&\text{ with }&\mat{C}_{\rvec{p}}^{\rvec{w}}=\diag[.1 \quad 40\mat{I}_2]
\end{eqnarray}
 MEM-EKF* (1) has less process noise  on the shape variables, i.e., only slight changes in the lengths of the semi-axes are possible. MEM-EKF* (2) has rather high process noise on the shape variables such that larger changes are possible.

The root mean squared Gaussian Wasserstein distance is given in Fig.~\ref{fig:turn_err}. 
As expected, the tracking performance improves significantly with turn rate estimation and a constant turn motion model. 
MEM-EKF* (1) has the least estimation error, as size changes are constrained using small process noise correspondingly.
However, note that, compared to random matrix approaches, MEM-EKF* involves more parameters. 
This is an advantage as well as a disadvantage. Its performance is sensitive to the  choice of the parameters, e.g.,  the results of MEM-EKF* (2) are worse as the used parameters  do not fit.

\section{Conclusion}\label{sec:conclusion}
Extended object tracking is challenging -- especially for the case of measurements that are scattered on the surface of the object.
As the underlying estimation problem is highly nonlinear, closed-form solutions are rarely available.

In this work, we introduced a new closed-form tracker called MEM-EKF* for the orientation and axes-lengths of an elliptical extended object. For this purpose, an explicit measurement equation is formulated via a multiplicative error. 
A problem-tailored combination of analytic moment calculation and linearization techniques then allows to derive a Kalman filter-based measurement update.
A major benefit of our method is that it provides an intuitive parameterization of an ellipse, which allows for directly modeling relevant motion models.
Furthermore, the full joint covariance of the ellipse parameters is available.
The closed-form formulas are compact, i.e., they are not significantly more complex than the standard Kalman filter formulas.

In the future, we will investigate extensions of the approach, e.g., for non-elliptical shapes or known extent \cite{Fowdur2019} and it will be embedded into multi-object trackers such as the extended target PHD filter \cite{Yang2019,Granstrom2012}.

\section*{Acknowledgment}
This work was supported by the German Research Foundation (DFG)
under grant BA 5160/1-1.

\begin{appendices}
\section{Derivations for the Joint Moments}
\subsection{Equation \eqref{eq:cov_2}}
\label{app:cov_2}
\label{app:cov_2}
As  $\rvec{h}$ is zero-mean,\footnote{For the sake of compactness, we omit the measurement index $(i)$ and time index $k$ }  for $m,n = 1,2$, we have
{\small
\begin{equation}
\cov{\vect{\rvec{h}^{\tp}\widehat{\mat{J_1}}\\\rvec{h}^{\tp}\widehat{\mat{J_2}}}\rvec{p},\vect{\rvec{h}^{\tp}\widehat{\mat{J_1}}\\\rvec{h}^{\tp}\widehat{\mat{J_2}}}\rvec{p}}
=[\epsilon_{mn}]\enspace,
\end{equation}
}
with
\begin{eqnarray}\label{eq:eps_1}
\epsilon_{mn}&=&\expect{\rvec{h}^{\tp}\widehat{\mat{J_m}}\rvec{p}\rvec{p}^{\tp}\widehat{\mat{J_n}}^{\tp}\rvec{h}},\\\label{eq:eps_2}
 &=& \expect{\trace{\rvec{h}^{\tp}\widehat{\mat{J_m}}\rvec{p}\rvec{p}^{\tp}\widehat{\mat{J_n}}^{\tp}\rvec{h}}},\\\label{eq:eps_3}
&=&\expect{\trace{\rvec{p}\rvec{p}^{\tp}\widehat{\mat{J_n}}^{\tp}\rvec{h}\rvec{h}^{\tp}\widehat{\mat{J_m}}}},\\\label{eq:eps_4}
&=&\trace{\expect{\rvec{p}\rvec{p}^{\tp}\widehat{\mat{J_n}}^{\tp}\rvec{h}\rvec{h}^{\tp}\widehat{\mat{J_m}}}},\\\label{eq:eps_5}
&=&\trace{\mat{C}^{\rvec{p}}\widehat{\mat{J_n}}^{\tp}\mat{C}^{\rvec{h}}\widehat{\mat{J_m}}}.\label{eq:eps_6}
\end{eqnarray}
Equation \eqref{eq:eps_2} follows from the fact that $\rvec{h}^{\tp}\widehat{\mat{J_m}}\rvec{p}\rvec{p}^{\tp}\widehat{\mat{J_n}}^{\tp}\rvec{h}$ is $1\times 1$.
As trace is invariant under cyclical permutations, we have \eqref{eq:eps_3}.
Equation \eqref{eq:eps_4} follows from the property that trace is a linear operator and can commute with expectation. 
Equation \eqref{eq:eps_5} follows form the independence between $\rvec{h}$ and $\rvec{p}$.
\subsection{Pseudo-measurement Covariance}\label{app:pseudo-cov}
To calculate the  covariance of the, pseudo-measurement we need the fourth centralized moments of original measurement, which can be calculated using Isserlis's theorem \cite{Isserlis1918} or Wick's theorem \cite{Wick1950}.
Given a measurement $\rvec{y}=\vect{y_1 & y_2}^{\tp}$,
the corresponding pseudo-measurement  is
\begin{equation}
\vect{\rvec{Y}_1\\\rvec{Y}_2\\\rvec{Y}_3} = \vect{(y_1-\bar{y}_1)^2\\(y_2-\bar{y}_2)^2\\(y_1-\bar{y}_1)(y_2-\bar{y}_2)}\enspace.
\end{equation}
From  Isserlis' theorem, we get
\begin{eqnarray}\label{eq:e_y1^2}
\expect{(\rvec{Y}_{1})^2}&=&3c_{11}^2\enspace,\\
\expect{(\rvec{Y}_{2})^2}&=&3c_{22}^2\enspace,\\
\expect{\rvec{Y}_{3}\rvec{Y}_{1}}&=&3c_{11}c_{12}\enspace,\\
\expect{\rvec{Y}_{3}\rvec{Y}_{2}}&=&3c_{22}c_{12}\enspace,\\
\expect{(\rvec{Y}_{3})^2}&=&\expect{\rvec{Y}_1\rvec{Y}_2}=c_{11}c_{22}+2c_{12}^2\enspace, 
\end{eqnarray}
where $c_{mn}$ denotes $\expect{(y_m-\bar{y}_m)(y_n-\bar{y}_n)}$ for  $m,n \in\{1,2\}$.
Based on the results above, the calculation of $mn$-th entry of pseudo-measurement covariance matrix  simply follows
\begin{equation}
\cov{\rvec{Y}_m,\rvec{Y}_n} = \expect{\rvec{Y}_m\rvec{Y}_n}- \expect{\rvec{Y}_m}\expect{\rvec{Y}_n}\enspace,
\end{equation}
for $m,n \in \{1,2,3\}$.
After a few further calculations, we get covariance of pseudo-measurement as in \eqref{eq:pseudo_cov}. 

\subsubsection{Linearization of the Pseudo-measurement Equation}
\label{app:jacobian}
 
Let $\mat{S_1}$ and $\mat{S_2}$ denote the first and second row of matrix $\mat{S}$. 
Similarly, $\mat{H_1}$ and $\mat{H_2}$ refer to the first and second row of $\mat{H}$.
Accordingly, the pseudo-measurement equation \eqref{eq:pseudo_meas} is rewritten as  (time  and measurement indices are omitted)
{{ 
\begin{equation}\label{eq:pseudo_meas_nok} 
g(\rvec{r},\rvec{p})=\vect{\left(\mat{H_1}\rvec{r}+\mat{S_1}\rvec{h}+v_{1}-\bar{y}_1\right)^2\\\left(\mat{H_2}\rvec{r}+\mat{S_2}\rvec{h}+v_{2}-\bar{y}_{2}\right)^2\\\left(\mat{H_1}\rvec{r}+\mat{S_1}\rvec{h}+v_1-\bar{y}_1\right)\left(\mat{H_2}\rvec{r}+\mat{S_2}\rvec{h}+v_{2}-\bar{y}_2\right)}
\end{equation}
}}
The cross-covariance of the pseudo-measurement and shape parameters are approximated using
\begin{eqnarray}
\mat{C}^{\rvec{p}\rvec{Y}} &=& \cov{\left.\frac{\partial g}{\partial \rvec{p}}\right|_{\rvec{p}=\hat{\rvec{p}}}(\rvec{p}-\hat{\rvec{p}}),\quad\rvec{p}}\\
&=& \mat{C}^{\rvec{p}}\left(\underbrace{\expect{\left.\frac{\partial g}{\partial \rvec{p}}\right|_{\rvec{p}=\hat{\rvec{p}}}}}_{\widehat{\mat{M}}}\right)^{\tp}
\end{eqnarray}

After applying the chain rule,  $\frac{\partial g}{\partial \rvec{p}}$ equals
{\footnotesize
\begin{equation}\label{eq:partila_g_p}
 \vect{2\left(\mat{H_1}\rvec{r}+\mat{S_1}\rvec{h}+v_{1}-\bar{y}_1\right)\rvec{h}^{\tp}\mat{J_1}\\
											2\left(\mat{H_2}\rvec{r}+\mat{S_2}\rvec{h}+v_{2}-\bar{y}_{2}\right)\rvec{h}^{\tp}\mat{J_2}\\
										\left(\mat{H_1}\rvec{r}+\mat{S_1}\rvec{h}+v_{1}-\bar{y}_1\right)\rvec{h}^{\tp}\mat{J}_{2}+\left(\mat{H_2}\rvec{r}+\mat{S_2}\rvec{h}+v_{2}-\bar{y}_{2}\right)\rvec{h}^{\tp}\mat{J}_{1}}
\end{equation}
}
with $\mat{J_1}$ and $\mat{J_2}$ are given in \eqref{eq:J1} and \eqref{eq:J2}.  
Evaluating the first row of \eqref{eq:partila_g_p} at $\hat{\rvec{p}}$, we have
\begin{equation}\label{eq:r1s1}
2(\mat{H_1}\rvec{r}-\bar{y}_1)\rvec{h}^{\tp}\widehat{\mat{J_1}}+2\widehat{\mat{S_1}}\rvec{h}\rvec{h}^{\tp}\widehat{\mat{J_1}}+2v_1\rvec{h}^{\tp}\widehat{\mat{J_1}}
\end{equation}
Taking the expectation of \eqref{eq:r1s1} gives us
\begin{equation}\label{eq:r1s2}
2\widehat{\mat{S_1}}\mat{C}^{\rvec{h}}\widehat{\mat{J_1}}
\end{equation}
After a  similar derivation for the second and third row of \eqref{eq:partila_g_p}, we get

{
\begin{equation}\label{eq:shape_jacobian}
\widehat{\mat{M}}=
			\vect{2\widehat{\mat{S_1}}\mat{C}^{\rvec{h}}\widehat{\mat{J_1}}\\
					2\widehat{\mat{S_2}}\mat{C}^{\rvec{h}}\widehat{\mat{J_2}}\\
					\widehat{\mat{S_1}}\mat{C}^{\rvec{h}}\widehat{\mat{J_2}}+\widehat{\mat{S_2}}\mat{C}^{\rvec{h}}\widehat{\mat{J_1}}}\enspace.
\end{equation}
}

\end{appendices}
\bibliographystyle{IEEEtran}

\bibliography{references.bib}

\end{document}